\documentclass[runningheads]{llncs}
\usepackage{graphicx}
\usepackage{amssymb}
\usepackage{amsmath}
\begin{document}
\title{De Novo Molecular Generation with Stacked Adversarial Model}
%
%
\author{Yuansan Liu\inst{1}\orcidID{0000-0003-3990-662X} \and
James Bailey\inst{1}\orcidID{0000-0002-3769-3811}}
\authorrunning{Yuansan et al.}
%
\institute{The University of Melbourne, Melbourne, VIC, Australia \\
\email{yuansanl@student.unimelb.edu.au,baileyj@unimelb.edu.au}}
\maketitle              
\begin{abstract}
Generating novel drug molecules with desired biological properties is a time consuming and complex task. Conditional generative adversarial models have recently been proposed as promising approaches for de novo drug design. In this paper, we propose a new generative model which extends an existing adversarial autoencoder (AAE) based model by stacking two models together. Our stacked approach generates more valid molecules, as well as molecules that are more similar to known drugs. We break down this challenging task into two sub-problems. A first stage model to learn primitive features from the molecules and gene expression data. A second stage model then takes these features to learn properties of the molecules and refine more valid molecules. Experiments and comparison to baseline methods on the LINCS L1000 dataset demonstrate that our proposed model has promising performance for molecular generation.
\keywords{De novo \and Molecule generation \and Adversarial autoencoder \and Stacking.}
\end{abstract}
\section{Introduction}
  During recent years, there has been great pressure on the pharmaceutical industry's productivity. To address this challenge, machine learning (ML) is being widely adopted in the biomedicine area including drug generation \cite{mamo}. For drug generation, researchers have focused mainly on conditional generative models, especially adversarial models like Generative Adversarial Networks (GANs) and Adversarial Autoencoders (AAE). Different from normal ones, a conditional generative model can learn a conditional distribution of a molecular structure with given properties, in order to generate molecules with specific properties. The adversarial training procedure facilitates the model to automatically learn the rules of valid structures and map them to latent representations. With such approaches, the range of candidate compounds can be narrowed down and ML generated molecule structures can be more valid.
  
  De novo molecular generation is an iterative process which designs new molecules using the structures of receptor proteins. Several recent works have studied such generation with desired bioactivity properties by using generative adversarial models, including \cite{fphar} which proposed a new AAE based model named the Bidirectional Adversarial Autoencoder(BiAAE), our main contribution in this paper is an enhancement of it, where we show how to effectively adapt the stacking method used in \cite{stack} to a new stacking method that works well for the BiAAE. By implementing this new stacking method, performance improves substantially, especially in the validity of the generation and the similarity with known compounds. Thus the generations have more likelihood to be biologically meaningful, potentially making it possible to increase the productivity of drug design.

  This paper is organized as follows: Section \ref{rw} presents a brief literature review. In section \ref{sbiaae} we will present our proposed model. Section \ref{exp} outlines experiments to evaluate our new model and compare with baselines. Finally, Section \ref{fi} concludes the paper.
  
\section{Related Works} \label{rw}
  \subsection{Generative Adversarial Models}
   Deep learning technologies have been applied in many areas over the last decade, among them, deep generative models have shown superior performance in image and text processing. Variational autoencoders (VAE) use a Bayesian method to learn the latent representations in order to turn the classic autoencoders into generative models \cite{kingma}, generative adversarial networks(GAN) use an adversarial approach in the training procedure to shape the output distribution \cite{gan}, adversarial autoencoders(AAE) combine the previous two models together \cite{makhzani}. There are also conditional extensions from these models: a conditional GAN (CGAN) concatenates label information into the generator's input noise to make GANs capable of more complex generation tasks \cite{mirza}, to improve the quality of generation, some researchers change the structure of the basic model: StackGAN stacked two GANs together to produce realistic images \cite{stack}.

  \subsection{Generative Models for Drug Discovery}
   Successes in machine learning for graphs and text has led to new applications of generative models in the drug discovery area. Initially, several works investigated VAE like model usage on generating simplified molecular-input line-entry system (SMILES) \cite{smile1} representation of molecules, the first of them being CharacterVAE \cite{charvae}. Apart from VAEs, GANs have also been used for this task, LatentGAN \cite{latentgan} combined an autoencoder and GAN for the de novo drug design task. In recent years, conditional generative models have been used to provide guidance for molecular generation in order to generate molecules with desired properties. In \cite{zhangli} the authors proposed a conditional graph generative model to manage de novo molecular design, \cite{limj} developed a generative model based on conditional VAE (CVAE) for molecular generation, \cite{masuda} conditioned a 3D binding pocket on the CVAE to generate 3D molecular structures for the first time.

   To make the generated molecules biologically meaningful, gene expression data, which proved to be useful in identifying novel active molecules \cite{dewolf}, roughly speaking, gene expression data records the activity (or change of activity) of genes within a cell. Researchers have thus started to condition generative models with gene expression data, e.g. in \cite{lucio}, a conditional WGAN is used for de novo generation of hit-like molecules from gene expression signatures. 
   
\begin{figure}[h]
     \centering
     \includegraphics[scale=0.4]{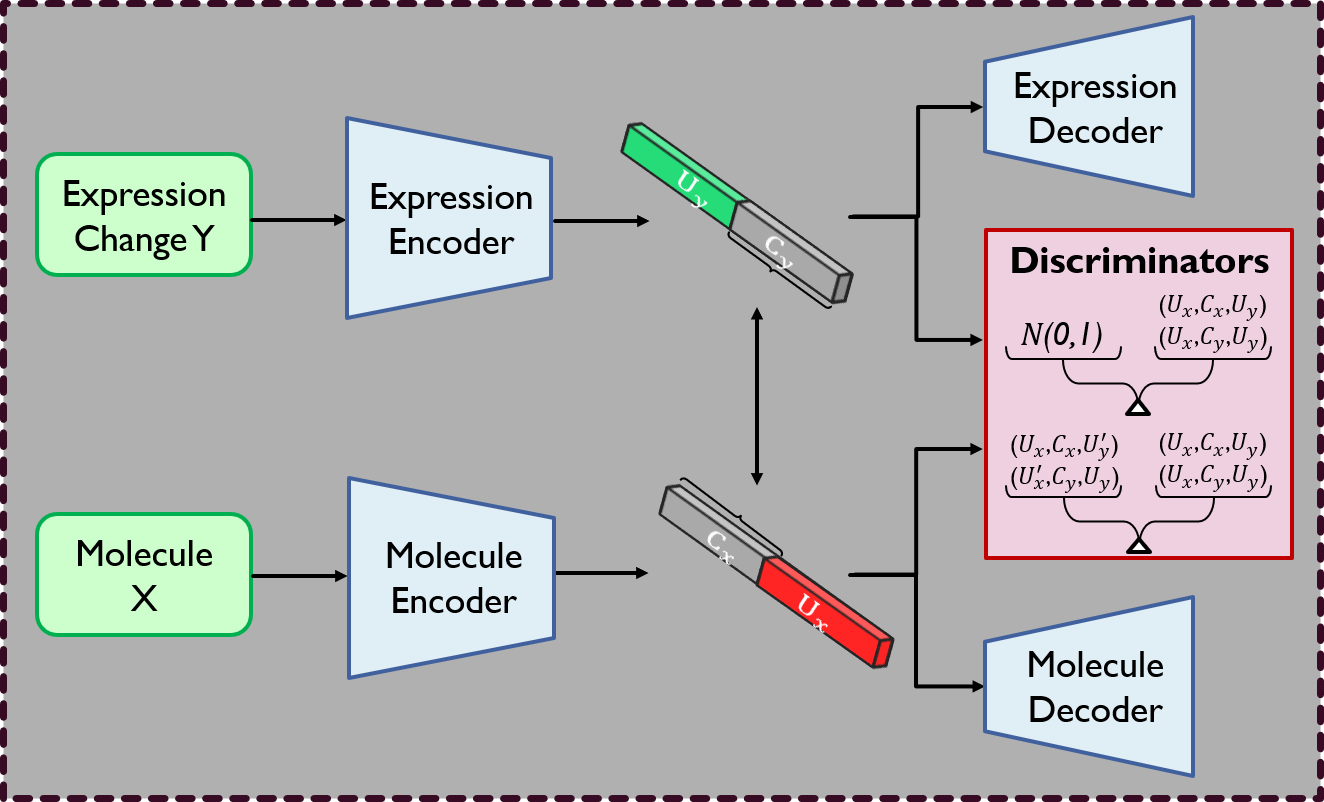}
     \caption{BiAAE Architecture}
     \label{f1}
   \end{figure}

   \cite{fphar} proposed an AAE based model BiAAE, which learns shared features between drug molecules and gene expression changes, to generate molecules with desired transcriptome changes. The latent codes of molecule $x$ and gene expression change $y$ are divided into unique part $u$ and common part $c$, $u$ represents the information that could be extracted only from $x$ or $y$, and $c$ is the information present in both $x$ and $y$, hence, $u$ is independent of $c$. To generate new synthetic data, the model samples unique parts $u_x, u_y$ and common parts $c$ independently from posterior distributions $G_x(x|c, u_x), G_y(y|c, u_y)$. Inference networks are used to predict latent codes: $u_x, u_y, c$ and train the model: $E_x(u_x|x), E_y(u_y|y), E_x(c|x)=E_y(c|y)=E(c|x, y)$. The architecture of the BiAAE is shown in the Figure \ref{f1}, deterministic encoders $E_x, E_y$ are used to hypothesize the latent representations of input data, and two deterministic decoders $G_x, G_y$ rebuild $x, y$ from their latent representations. Our model extended this work and achieved a better result, we will discuss it from next section.
   
\section{Stacked Bidirectional Adversarial Autoencoder} \label{sbiaae}

  It has been shown that stacking two or more models together can facilitate generating better synthetic samples in image generation task using GAN. In molecular generation, we can also expect for a similar effect.

  The targets of the adversarial training procedure are different between GAN and AAE. In GAN, the adversarial procedure is implemented over the generated samples, whereas in AAE, it is implemented over latent codes. This makes the stacking method of StackGAN inappropriate for the BiAAE. To solve this problem, we modify the original method: instead of the final output of the first stage, we use latent codes of condition $y$ in the first stage, together with one more learning of $x$, we get latent codes for the second stage. Specifically, the second stage does not have an expression encoder, the latent codes of $y$ are the ones generated by the first stage. For molecule data, we used the same learning procedure as the first stage to let it learn from molecules again to refine the learning outcome of the first stage, which is the common part of the $x,y$. With these modifications, we can now stack the BiAAE together to form the Stacked BiAAE (SBiAAE).
   
   \begin{figure}[ht]
     \centering
     \includegraphics[scale=0.4]{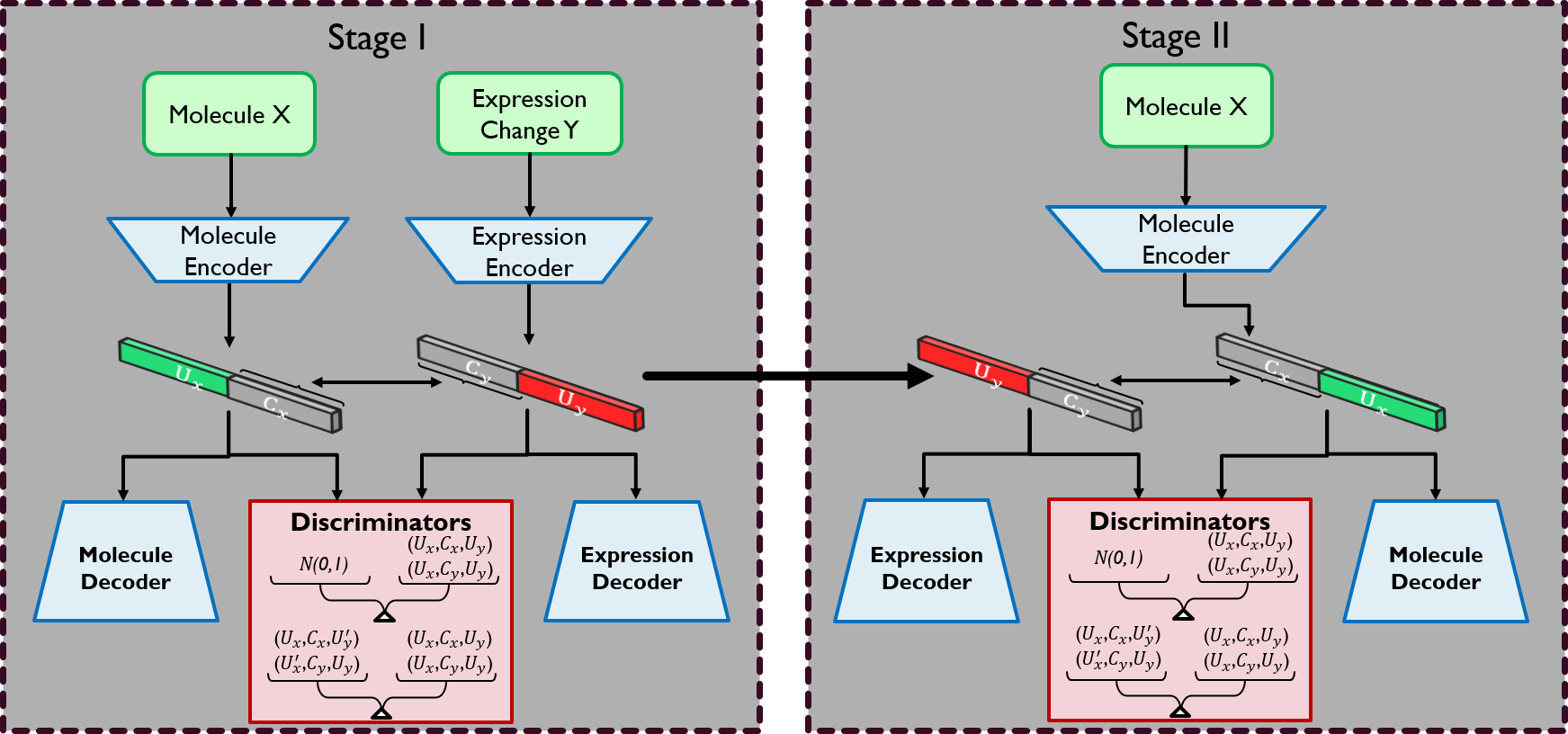}
     \caption{SBiAAE Architecture}
     \label{f2}
   \end{figure}

  Figure \ref{f2} shows the architecture of the SBiAAE, it can be seen that the stage one is the same as BiAAE and stage two is slightly different as it does not have a expression encoder but uses first stage's $y$ latent codes instead. Let $E_{ix,y}, G_{ix,y}, D_i, z_{ix,y}=(u_{ix,y}, c_{ix,y})$, where $i=1,2$, denote each part in the $i^{th}$ stage, we can get:$(u_{2x}, c_{2x}) \sim E_{2x}(x), x = G_{2x}(u_{2x}, c_{2x}), y = G_{2y}(u_{1y}, c_{1y})$, and the objective function of the stacked model has two stage, the first stage contains four parts:
   
   \begin{itemize}
     \item \textbf{Shared Loss}: ensures the common parts extract from $x$ and $y$ close to each other.
     \begin{equation}
       \mathcal{L}_{1share} = \mathbb{E}_{x,y \sim p_{d}(x, y)} \| c_{1x} - c_{1y} \|_{2}^{2} \label{ls}
     \end{equation}
     \item \textbf{Reconstruction Loss}: ensures the decoders can reconstruct the $x, y$ from their latent codes: $u_x, c_x, u_y, c_y$. Here the model added a cross-reconstruction loss, the decoders also used the common part from another one, for example, $(u_x, c_y)$ for $G_x$.
     \begin{align}
       \mathcal{L}_{rec}^x &= \mathbb{E}_{x\sim p_{d}(x)}l_{rec}^x(x, G_{1x}(u_{1x}, c_{1x})) + \mathbb{E}_{x,y \sim p_{d}(x,y)}l_{rec}^x(x, G_{1x}(u_{1x}, c_{1y})) \label{lrx}   \\
       \mathcal{L}_{rec}^y &= \mathbb{E}_{y\sim p_{d}(y)}l_{rec}^y(y, G_{1y}(u_{1y}, c_{1y})) + \mathbb{E}_{x,y \sim p_{d}(x,y)}l_{rec}^y(y, G_{1y}(u_{1y}, c_{1x})) \label{lry}
     \end{align}

     where $l_{rec}^{x,y}$  are similarity measures between original and reconstruct data.

     \item \textbf{Adversarial Loss}: uses a Jensen-Shannon Divergence(JSD) to guide the latent codes $z_{x,y} = (c_{x,y}, u_{x,y})$ to match an arbitrary prior distribution $\mathcal{N}(0,1)$, and encourage distributions $p(u_{x,y}), p(c_{x,y})$ to be independent.
     \begin{align}
       \mathcal{L}_{1adv} =& \mathbb{E}_{u_{1x}', c_{1x}', u_{1y}'\sim p(u_{1x}, c_{1x}, u_{1y})}logD(u_{1x}', c_{1x}', u_{1y}') \nonumber \\
        & +\mathbb{E}_{u_{1x}', c_{1y}', u_{1y}'\sim p(u_{1x}, c_{1y}, u_{1y})}logD_1(u_{1x}', c_{1y}', u_{1y}') \nonumber \\
        &+ \mathbb{E}_{x,y\sim p_{d}(x,y)}log(1-D_1(u_{1x},c_{1x},u_{1y})) \nonumber \\ 
        &+ \mathbb{E}_{x,y\sim p_{d}(x,y)}log(1-D_1(u_{1y}, c_{1y}, u_{1x}))  \label{ladv}
     \end{align}

     \item \textbf{Independence Loss}: explicitly encourages the independence of $u_x$ from $(u_y, c_y)$ and $u_y$ from $(u_x, c_x)$.
     \begin{align}
       \mathcal{L}_{ind} &= \mathbb{E}_{x,y\sim p_{d}(x,y)}\mathbb{E}_{y'\sim p_{d}(y)} [logD_1(u_{1x}, c_{1x}, u_{1y})+log(1-D_1(u_{1x}, c_{1x}, u_{1y}'))] \nonumber \\
       & + \mathbb{E}_{x,y\sim p_{d}(x,y)}\mathbb{E}_{x'\sim p_{d}(x)} [logD_1(u_{1x}, c_{1y}, u_{1y})+log(1-D(u_{1x}', c_{1y}, u_{1y}))]  \label{lind}
     \end{align}
     where $u_{1x}', u_{1y}'$ are obtained by shuffling $u_{1x}, u_{1y}$ in each batch.
   \end{itemize}

  The optimization problem is a minimax problem over the combination of (\ref{ls}) to (\ref{lind}):
   $$\min_{E_{1x,y}, G_{1x,y}} \max_{D_1} \lambda_1 \mathcal{L}_{share} + \lambda_2 \mathcal{L}_{rec}^x + \lambda_3 \mathcal{L}_{rec}^y + \mathcal{L}_{adv} + \mathcal{L}_{ind}$$
   The hyperparameters $\lambda_{1,2,3}$ are used to balance different objectives.
   
  Similar to the first stage, the objective functions of the second stage model are also divided into four parts:
   \begin{itemize}
       \item \textbf{Shared Loss}:
       \begin{equation}
           \mathcal{L}_{share} = \mathbb{E}_{x,y\sim p_{d}(x,y)}\|c_{2x}- c_{1y}\|_2^2 \label{ls2}
       \end{equation}
       \item \textbf{Reconstruction Loss}:
       \begin{align}
           \mathcal{L}_{rec}^x = & \mathbb{E}_{x\sim p_{d}(x)}l_{rec}^x(x, G_{2x}(u_{2x}, c_{2x})) +
       \mathbb{E}_{x,y \sim p_{d}(x,y)}l_{rec}^x(x, G_{2x}(u_{2x}, c_{1y})) \label{lrx2} \\
       \mathcal{L}_{rec}^y = & \mathbb{E}_{y\sim p_{d}(y)}l_{rec}^y(y, G_{2y}(u_{1y}, c_{1y})) +
         \mathbb{E}_{x,y \sim p_{d}(x,y)}l_{rec}^y(y, G_{2y}(u_{1y}, c_{2x})) \label{lry2}
       \end{align}
       \item \textbf{Adversarial Loss}:
       \begin{align}
            \mathcal{L}_{adv} = &\mathbb{E}_{u_{2x}', c_{2x}', u_{1y}' \sim p(u_{2x}, c_{2x}, u_{1y})}logD_2(u_{2x}', c_{2x}', u_{1y}') \nonumber \\
         &+ \mathbb{E}_{u_{2x}', c_{1y}', u_{1y}' \sim p(u_{2x}, c_{1y}, u_{1y})}logD_2(u_{2x}', c_{1y}', u_{1y}') \nonumber \\
         &+ \mathbb{E}_{x,y \sim p_{d}(x,y)}log(1-D_2(u_{2x}, c_{2x}, u_{1y})) \nonumber \\
         &+ \mathbb{E}_{x,y \sim p_{d}(x,y)}log(1-D_2(u_{2x}, c_{1y}, u_{1y})) \label{ladv2}
       \end{align}
       \item \textbf{Independence Loss}:
       \begin{align}
           \mathcal{L}_{ind} &= \mathbb{E}_{x,y \sim p_{d}(x,y)}\mathbb{E}_{y' \sim p_{d}(y)}[logD_2(u_{2x}, c_{2x}, u_{1y})+log(1-D_2(u_{2x}, c_{2x}, u_{1y}'))] \nonumber \\
       & + \mathbb{E}_{x,y \sim p_{d}(x,y)}\mathbb{E}_{x' \sim p_{d}(x)}[logD_2(u_{2x}, c_{1y}, u_{1y})+log(1-D_2(u_{2x}', c_{1y}, u_{1y}))] \label{lind2}
       \end{align}
   \end{itemize}

  Combine the equations (\ref{ls2}) to (\ref{lind2}), we can get the optimization problem for the second stage:
   $$\min_{E_{2x,y}, G_{2x,y}}\max_{D_2} \lambda_1 \mathcal{L}_{2share} + \lambda_2 \mathcal{L}_{2rec}^x + \lambda_3 \mathcal{L}_{2rec}^y + \mathcal{L}_{2adv} + \mathcal{L}_{2ind}$$
   
   
\section{Experiments} \label{exp}
  The original BiAAE model and the modified SBiAAE model are evaluated on two different datasets. An experiment on Noisy MNIST dataset is used as a preliminary evaluation from which we can test if the model has the potential to improve the original model. We then run the experiment on the LINCS L1000 molecular database to test the model's performance when conditioned using gene expression profiles.

  \subsection{Dataset Description}
   \begin{itemize}
     \item \textbf{Noisy MNIST}:\\
      The Noisy MNIST dataset was generated using the MNIST dataset \cite{nmnist}. In general, the images pair $(x,y)$ contain same digit with different angles and $y$ has strong additive noise. Thus, the only common information in $(x, y)$ is the digit they contain, the joint distribution $p_{xy}(x, y)$ should represent this digit. 

    \item \textbf{LINCS L1000}:\\
     The dataset is from the Library of Integrated Network-based Cellular Signatures (LINCS) L1000 project \cite{lincs}. The dataset is organized by cell lines, times, doses, molecules, and perturbagens. For each cell, it lists the gene expressions before and after($ge_a$, $ge_b$) it reacts with the given molecule, so we can obtain the gene expression changes for a given molecule by subtracting these two expressions.
   \end{itemize}

  \subsection{Experimental Setup}
    \subsubsection{Noisy MNIST}
     The whole Noisy MNIST dataset was split into three parts: 50000 training samples, 10000 validation samples and 10000 testing samples. The batch size was set to 128 with a learning rate 0.0003. The input data $x,y$ was encoded to a 16-dimensional tensor with 12-dimensional $u_x, u_y$ and 4-dimensional $c$. The same encoders for both $x,y$, each consisting of two convolutional layers with a 0.2 dropout rate, and rectified linear unit (ReLU) as activation function, then followed by three fully connected layers, with LeakyReLU activation function and batch normalization. The decoder has two fully connected layers with the exponential linear unit (ELU) activation function, followed by three transposed convolutional layers. The discriminator is a multilayer perceptron (MLP) with hidden layers of 1024 and 512 neurons. Balance weight $\lambda_{1,2,3}$ are set to 0.1, 10, and 1 respectively. The second stage is the same as the first one but without an encoder for $y$. The optimizer for the model is Adam with $\beta_1 = 0.5, \beta_2 = 0.9$ for adversarial training and $\beta_1 = 0.99, 0.999$ for others.

    \subsubsection{Gene Expression}
     For LINCS 1000 dataset, the training set contains experiments characterized by the control($ge_a$) and perturbation($ge_b$) induced gene expression profiles for each cell line. The molecules were represented by SMILES strings. The dataset was preprocessed based on Lipinski's rule of five \cite{LIPINSKI} and its extensions.

     The molecular encoder for $X$ took a pretrained RNN encoder's two-layers gated recurrent unit (GRU) as an initial part, followed by an fine-tuned MLP as second part. The expression encoder for $Y=(\Delta ge, \eta)$ embedded $\Delta ge = ge_b - ge_a$ using a two-layers MLP and concatenated the resulting hidden codes with logarithm of the concentration value $\eta$, processed it to another MLP to get the final representation. The decoders for two objects were built symmetrically to the corresponding encoders. The second stage used the same structure except the expression encoder, which does not exist, and the latent codes of the gene expression $Y$ in this stage are the results of the first stage expression encoder.

  \subsection{Experimental Evaluations}
   \subsubsection{Metrics}
    \begin{itemize}
      \item \textbf{Accuracy}: \\
       Accuracy is mainly used for evaluation on the Noisy MNIST experiment, the ratio between correction generation and total number generated is the accuracy of the model.
      \item \textbf{Mutual Information(MI)}: \\
       We calculated the mutual information using MINE \cite{mine}, this score can represent the ability of the encoders to extract relevant information.
      \item \textbf{Validity}: \\
       Validity is the proportion between valid number and total number, it evaluates how often the model produces valid molecule.
      \item \textbf{Overall Similarity}: \\
       The similarity between generated molecules and their templates are calculated using Tanimoto similarity of their Extended Connectivity Fingerprints(ECFPs) \cite{ecfps}. $\frac{\sum T(A,B)}{N}, T(A,B) = \frac{A\cap B}{A+B-A\cap B}$. where $A, B$ are ECFPs for generated and template molecules, $T$ is Tanimoto similarity, $N$ is total number of valid generation.
      \item \textbf{Diversity}: \\
       $1-\frac{1}{N}\sum_{i, j>i} T(m_i, m_j)$ for $N$ pairs of valid generated molecules, normally $N=\binom{n}{2}$ where $n$ is total number of valid generation.
    \end{itemize}

  \subsubsection{Noisy MNIST Experiment}
\begin{figure}[ht]
    \centering
      \includegraphics[scale=0.3]{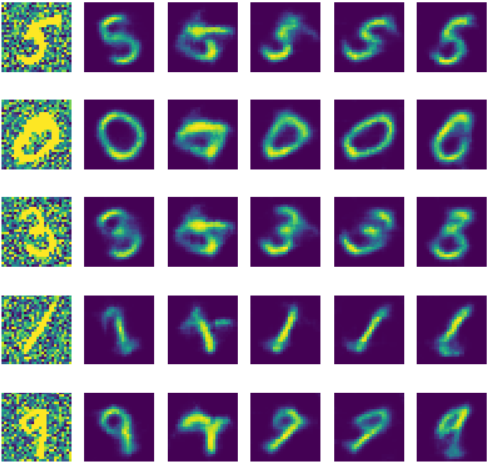}%
      \hspace{1cm}
      \includegraphics[scale=0.3]{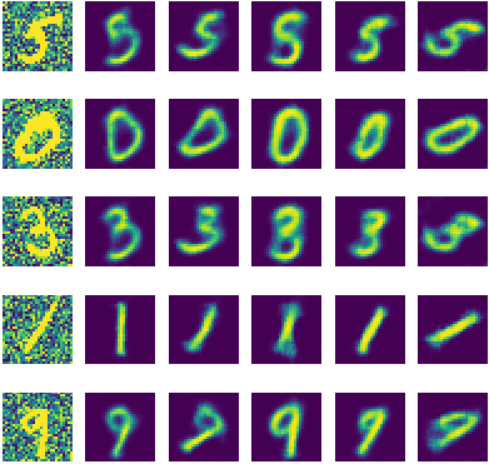}
    \caption{Generated figures by BiAAE and SBiAAE}
    \label{f3}
   \end{figure}

   The experiment run over Noisy MNIST dataset is intended to test model performance on common information extraction, this ability should be reflected by the accuracy and mutual information scores. We use $MI(z_x, c_y|y)$ for mutual information score, it represents how much the $c_y$ can indicate about $z_x$. The only common thing between them is $c_y$ itself, so, higher $MI$ score means the encoder produced more desirable latent codes. The quantitative results for several models run on the Noisy MNIST shown in the Table \ref{t1}, the UniAAE is the unidirectional version of BiAAE. SAAE is conditional AAE mentioned in \cite{makhzani}, we use $MI(z_x, c|y)$ since it cannot extract the common parts from the latent codes explicitly.
   \begin{table}[ht]
    \centering
    \caption{Noisy MNIST results}
    \begin{tabular}{cccc}
     \hline
     Model & Accuracy & MI($z_x, c_y|y$) & MI($z_x, c|y$)\\
     \hline
     Stacked BiAAE & \textbf{68.06} & \textbf{1.729} & - \\
     BiAAE & 61.34 & 1.432 & - \\
     UniAAE & 50.11 & 1.550 & - \\
     SAAE & 45.46 & - & \textbf{1.673} \\
     \hline
    \end{tabular}
     \label{t1}
   \end{table}
   Examples of some generated images from BiAAE(left) and SBiAAE(right) are shown in the Figure \ref{f3}. Comparing two figures, we can observe that samples generated by BiAAE still contain some wrong digits, whereas most of the SBiAAE generations are correct digits.

   From the results in Table \ref{t1} and examples in Figure \ref{f3}, we can see the SBiAAE did well in extracting digits from the image pairs as intended. Based on this performance, we can draw an initial conclusion that the stacked model is better at joint distribution learning for Noisy MNIST dataset, in next part we would run the experiment on gene expression dataset to check how well the model perform for our real task.

  \subsubsection{LINCS L1000 Experiment}
   The results from Noisy MNIST dataset gave us enough confidence to run the model on LINCS L1000 dataset.
   
   In modern chemistry, a pharmacophore is a part of a molecular structure which can be used to represent an abstraction of the molecular features and is capable of inferring the affected genes, since it ensures the optimal molecule interactions with a single specific biological target so that it would trigger (or block) its biological response \cite{gloss}.  This makes the latent representations of pharmacophore and the affected proteins similar and hence in our joint distribution learning, the common part between the given molecule $x$ and gene expression $y$ can be the pharmacophore in the molecule and affected protein in gene expressions.

   The mutual information used in this experiment is $MI(z_x, c_y|\Delta ge, \eta)$ which calculates the relevant information between molecular latent codes and common part of gene expressions extracted by the expression encoder. Together with overall similarity, they reveal the ability of the encoders for extracting common parts from molecules and gene expressions. The validity reflects how well the decoder learnt the rules of the molecular structure. As a de novo generation task, the model is also expected to keep a good diversity in generation while raising the similarity to the templates.

   The results of models for molecular generation are shown in Table \ref{t2}, in this part we only list two models: BiAAE and SBiAAE for direct comparison.
   \begin{table}[ht]
    \centering
    \caption{LINCS L1000 results}
    \begin{tabular}{ccccc}
     \hline
      Model & MI($x,s_y|\Delta ge, \eta$) & Validity & Overall Similarity & Diversity\\
      \hline
      BiAAE & 0.28 & 64.7\% & 0.247 & 0.85 \\
      SBiAAE & 0.29 & 69.1\% & 0.265 & 0.83 \\
      \hline
    \end{tabular}
    \label{t2}
   \end{table}

\begin{figure}[ht]
     \centering
     \includegraphics[scale=0.4]{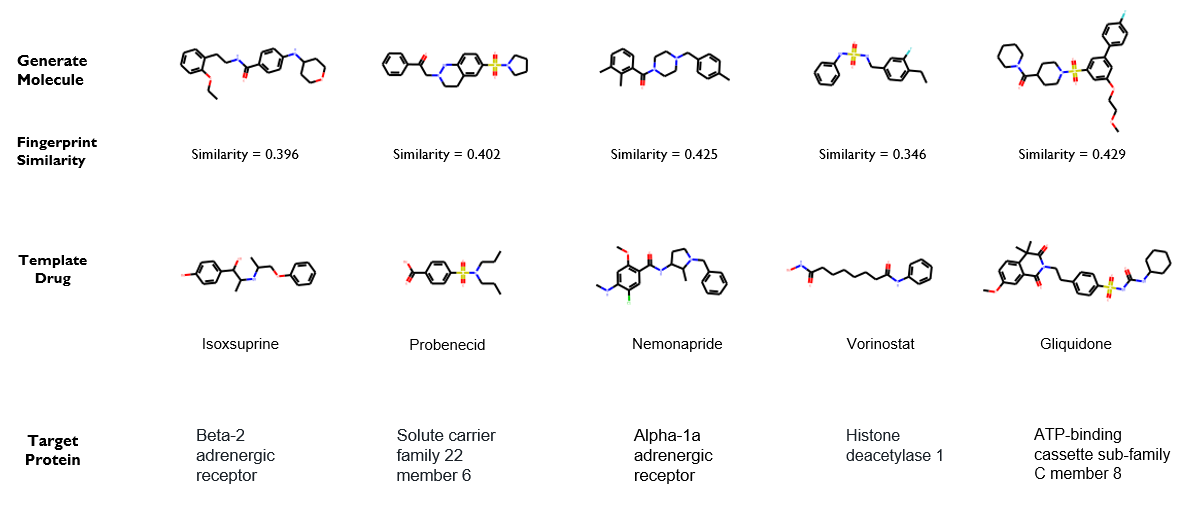}
     \caption{Generated molecules with their drug templates}
     \label{f4}
   \end{figure}

   Examples of generated molecules together with their templates, which are active drugs in real world, are shown in Figure \ref{f4}.
   Here we listed the fingerprint similarities between the generations and corresponding templates that are calculated based on their ECFPs, the target proteins are also listed for each generated molecule. The main intuition for this presentation is the models are expected to generate some biologically meaningful molecules, thus by presenting the generated molecule and its template drug, we can show that these generated molecules indeed do have bioactivity.  This is due to the similarity principle which states ``structurally similar molecules tend to have similar properties in both physicochemical and biological ones'' \cite{concept}, and the pharmacophore's application on defining the essential features of one or more molecules with the same biological activity.

   From Table \ref{t2} we can see the greatest improvement that our stacked model achieves is on generation validity. The stacking method allows the model to learn more from molecular data and, as intended, it learns more rules from the valid molecules and generates more valid molecules.

   For common part extraction, the SBiAAE does improve the performance a little. The mutual information scores indicate encoders from stacked model extracted the common part slightly better than the original model, a similar result is also shown in the Overall Similarity. These two scores suggest the SBiAAE extracts something extra from the molecules and the gene expression changes, and the resulting common parts of gene expression changes contain more relevant information about the molecules, which as previously mentioned could be the pharmacophore of the molecule.
   \begin{figure}[ht]
     \centering
     \includegraphics[scale=0.4]{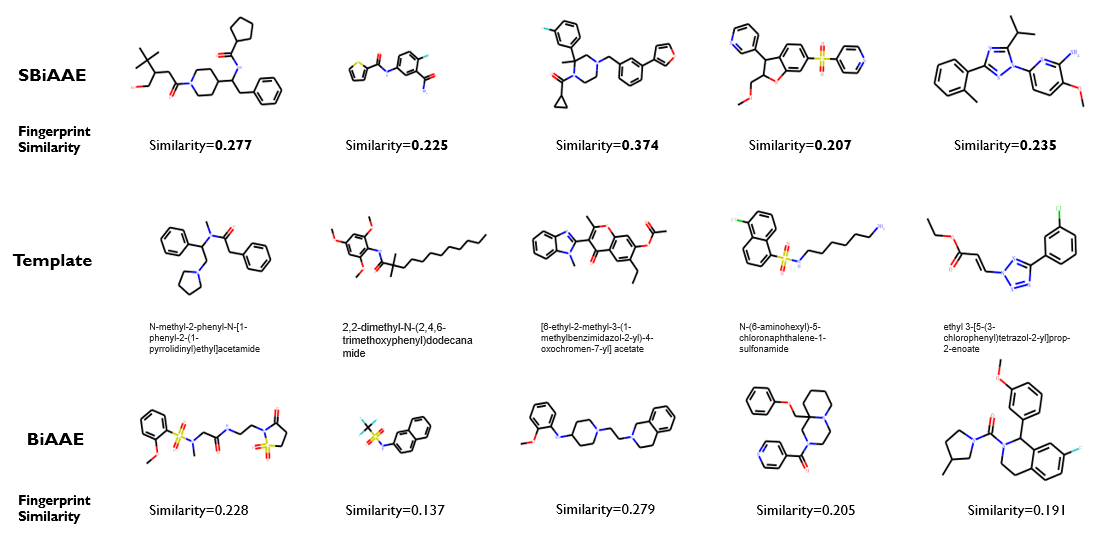}
     \caption{Generations of BiAAE and SBiAAE}
     \label{f5}
   \end{figure}
   In Figure \ref{f5} we present a comparison between two models' generations, and the templates they based on. For better comparison, we also listed the similarity for each molecule. Some of the generations from SBiAAE are much more similar to the templates like the second and third columns.

   In general, the stacked model SBiAAE improved the performance of the original BiAAE model, the most obvious improvement lies in the validity of the generated molecules. For common information extraction, the stacking method improved the original model, though it did not improve as much as for the Noisy MNIST experiment. Based on the generated samples, we can also say that the generated molecules might be biologically meaningful because of the improvements on the average similarity, the stacked model can have more chance to generate the bioactive molecules compared to the original model. Overall, for this dataset, we have evidence that SBiAAE performs better than BiAAE for de novo generation of desired molecules.
   
\section{Conclusion} \label{fi}

   To conclude, in this project, we studied an existing model for de novo molecular generation named BiAAE, and investigated the machine learning technologies and biochemical knowledge related to it. Then, we developed a new stacking method derived from StackGAN to make it appropriate for the joint distribution learning task in BiAAE, with this method we stacked an other model on the original BiAAE and improved the performance of BiAAE on a molecular generation task. The main contribution of our work is the modified stacking method which made it possible for the new SBiAAE model to be more reliable for generating molecule.

\bibliographystyle{splncs04}
\bibliography{ref}

\end{document}